# ON THE NUMERICAL SOLUTION OF A VARIABLE-COEFFICIENT BURGERS EQUATION ARISING IN GRANULAR SEGREGATION

I.C. Christov

School of Mechanical Engineering, Purdue University, West Lafayette, Indiana 47907, USA

e-mail: christov@purdue.edu

**Abstract.** We study a variable-coefficient Burgers equation arising in the modelling of segregation of dry bidisperse granular mixtures. The equation is subject to nonlinear boundary conditions for the particle flux. We construct a strongly implicit Crank–Nicolson type of numerical scheme for the latter equation. The scheme is benchmarked against a standard exact solution of kink type, showing second-order of accuracy and good discrete conservation properties. Two segregation problems considered in the literature are then solved and discussed. The first is the case of a linear kinetic stress profile, which renders the governing equation of constant-coefficient type, while the second is the case of a variable kinetic stress profile based on an expression fit to particle dynamics simulation data.

## 1. Introduction

In this short communication, we wish to solve numerically the following variable-coefficient Burgers partial differential equation originally due to Dolgunin and Ukolov [1] (see also the review by Gray et al. [2]):

$$\frac{\partial \phi}{\partial t} + \frac{\partial}{\partial y}[\phi(1-\phi)f(y)] - D\frac{\partial^2 \phi}{\partial y^2} = 0, \qquad y \in [-L, L], \qquad t > 0, \tag{1}$$

where $\phi = \phi(y,t)$ is the concentration of either large $\phi^l$ or small $\phi^s$ small particles, where $\phi^l + \phi^s = 1 \,\forall t, y$ since we restrict to binary granular mixtures. The function $f(y)$ (units of length/time) depends on the granular flow under consideration and is sometimes called "the percolation velocity." $D$ is the constant diffusivity (due to inter-particle collisions) of the flowing granular material. Recently, models of the type (1) have been extended to include mean flow variations in chutes [3] and the polydispersity of the granular material [4,5].

Equation (1) represents perhaps the simplest, yet quite accurate, model of the evolution of the concentration of particles in a granular flow. Two main effects are captured therein: (*i*) diffusion due to inter-particle collisions, which tends to "smooth out" concentration gradients, and (*ii*) segregation due to the ability of smaller particles to fall through the gaps of larger particles, which tends to "steepen" the concentration profiles by sifting small particles from large ones. These effects are, of course, conceptually identical to those encountered in the hydrodynamics context in which Burgers' equation was first considered (see, e.g., [6]). Note, however, that the segregation term "shuts off" for a mixture consisting of only one type of particle, i.e., $\phi = 0$ or $1$, *unlike* the hydrodynamic Burgers equation in which case the advective flux only vanishes at the zero state. On the other hand, this advective nonlinearity of (1) is *identical* to the one in the Lighthill–Whitham–Richards kinematic theory of traffic flow [6].

We are specifically interested in the case of $f(y) = (q/\rho)(\partial \sigma_{yy}^k/\partial y)$, where $q$ is sometime termed the "segregation coefficient" [7] and has units of time, $\rho$ is the average local density of the mixture, and $\sigma_{yy}^k$ is a component of the kinetic particle stress of the granular mixture [7,8]. Specific examples of $f(y)$ will be discussed in Sections 4 and 5 below.

Based on (1), we can define the particle flux $Q := \phi(1-\phi)f(y) - D\partial\phi/\partial y$. We assume, further, that $y = \pm L$ represents solid walls, hence there cannot be particle flux there: $Q(\pm L, t) = 0 \;\forall t$. Therefore, the appropriate boundary conditions (BCs) for (1) are

$$\phi(1-\phi)f(y) = D\frac{\partial \phi}{\partial y}, \qquad y = \pm L, \tag{2}$$

where we note that these BCs are, in fact, *nonlinear*. Equations (1) and (2) constitute the model put forth by Gray and Chugunov [9] for granular segregation and diffusion in a unidirectional flow, generalizing the model of Gray and Thornton [10], which neglected diffusion. Except for Section 3, we will consider one type of the initial condition for (1), specifically a well-mixed granular state, i.e., $\phi(y, 0) = \phi_0 \;\forall y \in [-L, L]$ for some initial concentration $0 < \phi_0 < 1$.

The remainder of this work is organized as follows. The main result, presented in Section 2, is the construction of a fully-implicit Crank–Nicolson-type scheme for (1) subject to (2). In Section 3, a particular exact kink-shaped solution for the constant-coefficient Burgers equation is derived, and the scheme is benchmarked against this solution. In Sections 4 and 5, numerical simulations are presented and discussed for two different cases of the function $f(y)$. Finally, in Section 6, conclusions are stated and future work is suggested.

## 2. A fully-implicit Crank–Nicolson scheme

We define the uniform spatial grid spacing and temporal step size as $\Delta y := 2L/(M-1)$ and $\Delta t := T/(K-1)$, respectively, for some integers $K \geq 1$ and $M \geq 1$. Hence, there are $M$ grid points in the domain and $K$ time steps are taken, and we have assumed a computational domain $(y, t) \in [-L, L] \times (0, T]$. Let the grid function $\Phi_j^n \approx \phi(y_j, t^n)$ represent the approximation to the exact solution $\phi$ on the discrete grid, where $y_j := -L + j\Delta x$ ($0 \leq j \leq M-1$) and $t^n := n\Delta t$ ($0 \leq n \leq K-1$). Following Strikwerda [11], we denote by $\delta_{x\pm}$ and $\delta_{t\pm}$ the spatial and temporal forward and backward finite difference operators, while $\delta_{x0}$ and $\delta_{t0}$ represent the spatial and temporal central difference operators, respectively.

To obtain desirable stability and discrete conservation properties [12,13], we construct a two-level "strongly-implicit" (i.e., implicit diffusion, implicit advection) scheme for (1) by using a first-order forward time discretization but taking every $\phi$ term to be the average of the grid functions at the current *and* next time step:

$$\delta_{t+}\Phi_j^n + \delta_{y0}\left\{\left(\frac{\Phi_j^{n+1} + \Phi_j^n}{2}\right)\left[1 - \left(\frac{\Phi_j^{n+1} + \Phi_j^n}{2}\right)\right]f(y_j)\right\} - D\delta_{x+}\delta_{x-}\left(\frac{\Phi_j^{n+1} + \Phi_j^n}{2}\right) = 0. \tag{3}$$

Hence, this is a scheme of the Crank–Nicolson type [14]. We have used second-order central spatial difference operators for the $y$ derivatives, thus the scheme is formally $O[(\Delta y)^2 + (\Delta t)^2]$ accurate for smooth solutions.

Equation (3) can be rearranged in a more standard form:

$$\left[\frac{\Delta t}{8\Delta y}(2 - \Phi_{j+1}^{n,k} - 2\Phi_{j+1}^n)f(y_{j+1}) - \frac{D\Delta t}{2(\Delta y)^2}\right]\Phi_{j+1}^{n,k+1} + \left[1 + \frac{D\Delta t}{(\Delta y)^2}\right]\Phi_j^{n,k+1}$$

$$+ \left[-\frac{\Delta t}{8\Delta y}(2 - \Phi_{j-1}^{n,k} - 2\Phi_{j-1}^n)f(y_{j-1}) - \frac{D\Delta t}{2(\Delta y)^2}\right]\Phi_{j-1}^{n,k+1}$$

$$= -\frac{\Delta t}{8\Delta y}\left[\Phi_{j+1}^n(2 - \Phi_{j+1}^n)f(y_{j+1}) - \Phi_{j-1}^n(2 - \Phi_{j-1}^n)f(y_{j-1})\right] + \Phi_j^n$$

$$+ \frac{D\Delta t}{2(\Delta y)^2}(\Phi_{j+1}^n - 2\Phi_j^n + \Phi_{j-1}^n), \tag{4}$$

where in anticipation of the next step in the procedure, we have replaced *some* terms on the left-hand side by either $\Phi^{n,k}$ or $\Phi^{n,k+1}$ ($j$ subscripts understood). By doing so, we have linearized the intrinsically *nonlinear* scheme (3) and set it up for a fixed-point iteration procedure. This is known "internal iterations" [12,13], due to Yanenko [15]. The iteration scheme (4) is initialized with the data from the last time step: $\Phi_j^{n,0} = \Phi_j^n \;\forall j$. Then, (4) becomes a tridiagonal linear system for $\Phi^{n,1}$ ($j$ subscripts understood). The linear system is solved by standard methods in MATLAB. The iteration continues until the (relative) termination criterion $\max_j |\Phi_j^{n,k^*} - \Phi_j^{n,k^*-1}| < 10^{-8} \max_j |\Phi_j^{n,k^*-1}|$ is met for some $k = k^*$. Then, the solution at the new time step has been found: $\Phi_j^{n+1} = \Phi_j^{n,k^*} \;\forall j$. Typically, only a few iterations are required at every time step to meet the termination criterion.

Next, the boundary conditions must be discretized. To maintain the overall second-order of accuracy of the scheme, we must use $O[(\Delta y)^2]$ finite-difference formulas to discretize (3). To this end, we employ the second-order-accurate forward/backward difference representations of the first derivative [11]. Then, upon linearizing the discrete versions of (2) to match the internal iteration scheme (4), we obtain:

$$\Phi_0^{n,k+1}(1 - \Phi_0^{n,k})f(y_0) = \frac{D}{2\Delta y}\left(-\Phi_2^{n,k+1} + 4\Phi_1^{n,k+1} - 3\Phi_0^{n,k+1}\right), \tag{5a}$$

$$\Phi_{M-1}^{n,k+1}(1 - \Phi_{M-1}^{n,k})f(y_{M-1}) = \frac{D}{2\Delta y}\left(3\Phi_{M-1}^{n,k+1} - 4\Phi_{M-2}^{n,k+1} + \Phi_{M-3}^{n,k+1}\right). \tag{5b}$$

Equations (5) can each be solved for $\Phi_0^{n,k+1}$ and $\Phi_{M-1}^{n,k+1}$, respectively, thereby closing the linear system (4) used for the internal iterations.

Finally, note that (1) can be integrated over $y \in [-L, L]$, to obtain $\frac{d}{dt}\int_{-L}^{+L} \phi\, dy = Q(-L, t) - Q(+L, t)$. By the no-flux BCs in (2), $\int_{-L}^{+L} \phi\, dy = const.$ Our numerical scheme (4) inherits this conservation property. As can be easily shown by a telescoping series argument, $\sum_{j=0}^{M-1} \Phi_j^{n+1} = \sum_{j=0}^{M-1} \Phi_j^n$.

## 3. Special solutions, benchmarking and estimated order of convergence of the scheme

Now, consider (1) with $f(y) = f = const.$ on $y \in (-\infty, +\infty)$ subject to asymptotic boundary conditions $\partial\phi/\partial y, \partial\phi/\partial t \to 0$ as $|y| \to \infty$. By standard methods [16], we can show there exists an explicit traveling wave solution:

$$\phi(y,t) = \psi(y - ct) = \frac{c-f}{2f}\left\{-1 + \tanh\left[\frac{(c-f)(y-y_0-ct)}{2D}\right]\right\}, \tag{6}$$

where $c$ is the (arbitrary) wave speed, and $y_0$ is another free parameter due to translation invariance of (1). The solution (6) has a clear physical interpretation: this is a transition segregation front connecting an equilibrium $\phi = 0$ or 1 to a segregated state $(f-c)/f$ (or $c/f$ for the reciprocal or, "anti," kink), which depends on the percolation velocity and the speed of the front. However, since $\phi$ is a concentration such that $0 \le \phi \le 1$, then for $f \gtrless 0$, we must have $c \gtrless 0$ to satisfy this requirement. Either way, it is clear that faster fronts leave "less" segregation behind. A comparison of the solution (6) and its numerically computed counterpart is shown in Fig. 1.

Now, we would like to use the exact solution (6) to benchmark the accuracy and to estimate the order of convergence of the proposed scheme (4). Notice, however, that the kink solution (6) only satisfies *homogeneous Neumann* BCs ($\partial\phi/\partial y \approx 0$ for $x = \pm L$, as long as $L$ is sufficiently large for the chosen $D, y_0, c, t$), hence the BCs (5) have to be changed to

$$0 = \frac{1}{2\Delta y}\left(-\Phi_2^{n,k+1} + 4\Phi_1^{n,k+1} - 3\Phi_0^{n,k+1}\right), \tag{7a}$$

$$0 = \frac{1}{2\Delta y}\left(3\Phi_{M-1}^{n,k+1} - 4\Phi_{M-2}^{n,k+1} + \Phi_{M-3}^{n,k+1}\right), \tag{7b}$$

for the purposes of the estimated order of convergence study.

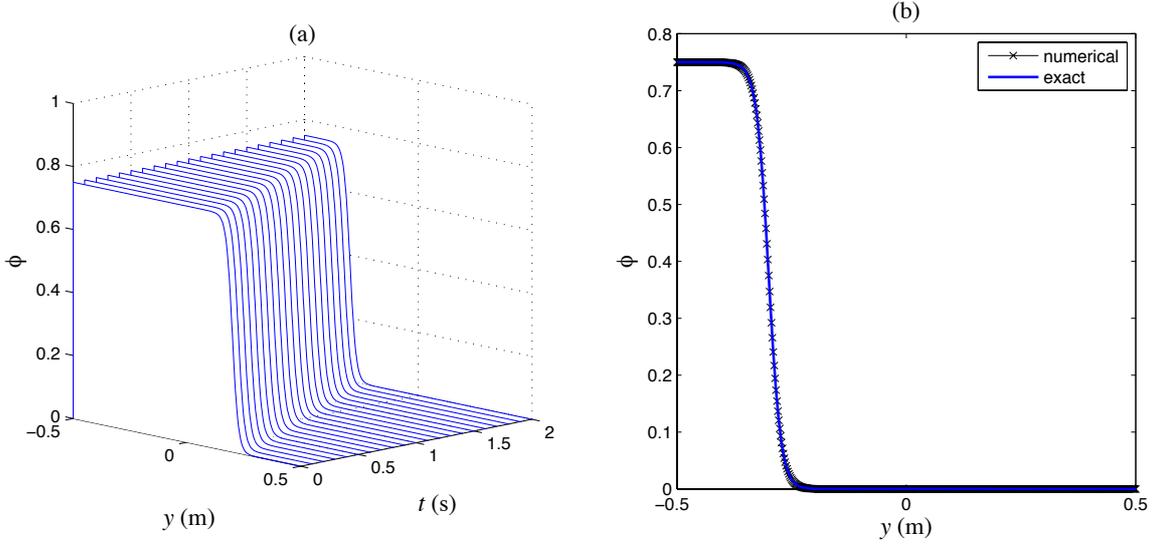

Figure 1. Comparison between the exact (6) and numerical kink solutions of the governing equation (1). (a) A "waterfall" plot of the numerical solution as a function of $y$ and $t$; (b) comparison between numerical and exact $\phi(y, t = 2\text{ s})$. The parameters for this case are $f = -1$, $D = 0.01\text{ m}^2/\text{s}$, $y_0 = 0.2\text{ m}$, $c = -0.25\text{ m/s}$, $\Delta t = 0.002\text{ s}$ and $\Delta y = 1\times 10^{-3}$ m on the domain with $L = 0.5$ m at $t = T = 2$ s ($\Rightarrow M = 2001, K = 501$).

The error between the exact and numerical solutions is computed at $t = T = 2$ s (as shown in Fig. 1(b)) in the usual max-norm, $\|\cdot\|_{L^\infty}$, and two Sobolev functional norms:

$$\|\cdot\|_{L^2} = \left(\int_{-L}^{+L} |\cdot|^2\, dy\right)^{1/2}, \tag{8a}$$

$$\|\cdot\|_{H^1} = \left(\int_{-L}^{+L} |\cdot|^2\, dy + \int_{-L}^{+L} \left|\frac{\partial(\cdot)}{\partial y}\right|^2 dy\right)^{1/2}. \tag{8b}$$

By halving $\Delta t$ and $\Delta y$ simultaneously, the estimated order of convergence in the three chosen norms can be calculated by standard formulas [11]. As the kink solution (6) is smooth ($C^\infty$, in fact), we observe second-order accuracy in all three norms, as shown in Fig. 2. The error is defined as the numerical solution minus the exact solution, at the grid points $y_j$.

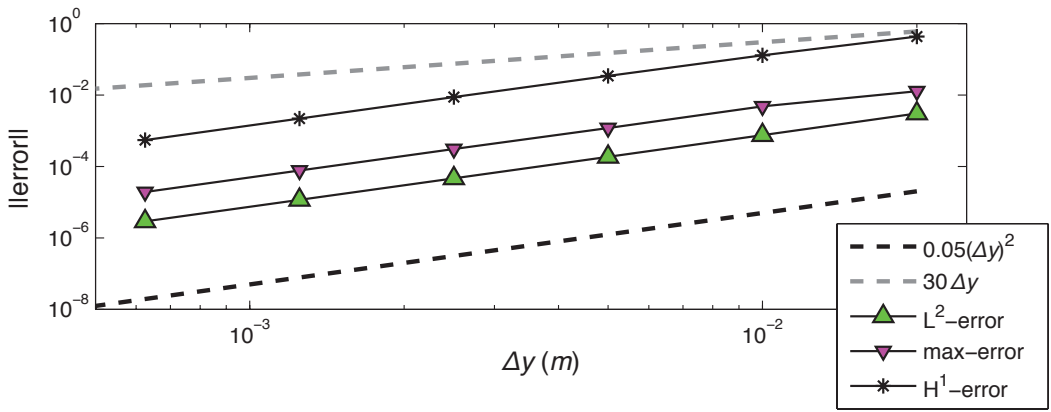

Figure 2. Grid refinement study based on the exact solution (6), with $f = -1$, $D = 0.01\text{ m}^2/\text{s}$, $y_0 = 0.2$ m and $c = -0.25$ m/s, for the scheme (4) with BCs (7) on a domain $L = 0.5$ m at the final time $T = 2$ s. Both $\Delta t$ and $\Delta y$ were halved simultaneously from initial values of $\Delta t = 0.02$ s, $\Delta y = 0.02$ m.

## 4. Simulations of granular segregation: Linear kinetic stress profile

If the kinetic stress profile is linear in $y$, then we can write $\partial \sigma^k_{yy}/\partial y = K$, for some constant $K$, and $f(y) = qK/\rho =: \hat{q}$. This case was considered by Gray and Chugunov [9] and an exact solution to the initial–boundary–value problem (IBVP) (1)–(2) was obtained via the Cole–Hopf transformation [6]. This elegant result relies entirely on the fact that the governing equation has constant coefficients in this special case. With our numerical scheme, we need not make such assumptions and can solve this case as we would any other.

For consistency with [9], we take $\hat{q} = -1$ m/s and $D = D_r |\hat{q}|(2L)$ ($D_r = 0, 0.02, 0.1, 0.5$ in [9]) and $T = \tilde{t}(2L/|\hat{q}|)$ ($\tilde{t} \approx 2$ in [9]). Without loss of generality, we normalize the domain to $2L = 1$ m. Of course, we cannot treat $D_r = 0$ since this is a singular limit. Instead, we choose $D_r = 0.002, 0.02, 0.1, 0.5$ and our numerical solutions for these values are shown in Fig. 3. In all cases, the no-flux BC at the walls drives segregation fronts into the core of the granular material (initially well mixed with $\phi_0 = 0.55$), which leads to the eventual segregation into two distinct phases for small $D_r$. For larger $D_r$, diffusion causes remixing, counteracting segregation.

These plots can be compared to left column of Fig. 6 in [9], which, of course, is the *exact* solution to the IBVP via the Cole–Hopf transformation. A close look reveals that our numerical results are visually indistinguishable from the exact solutions plotted in [9]. Furthermore, though we cannot take $D_r = 0$, our scheme remains stable and highly accurate as $D_r$ is made very small, and the segregation fronts become very steep (Fig. 3, top left plot).

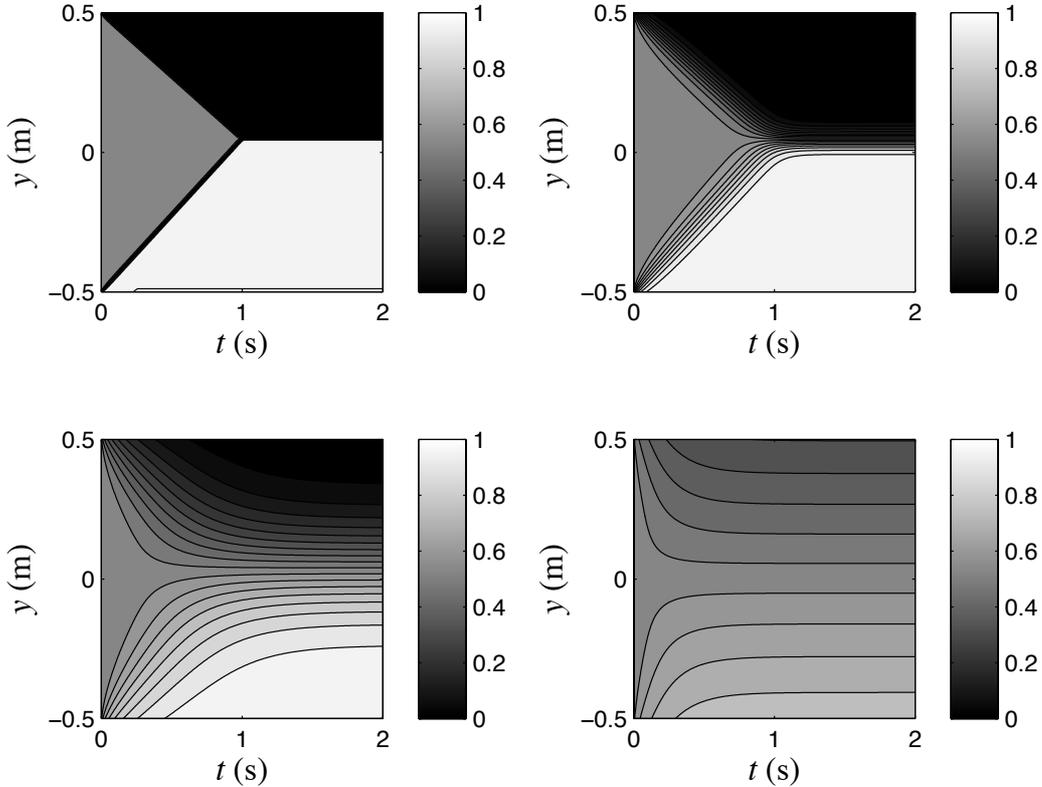

Figure 3. Contour plots of the spatiotemporal evolution of $\phi(y, t)$ (color bars denotes $\phi$ values for contours) as an initially uniform mixture segregates under a linear kinetic stress profile with $\Delta y = 2 \times 10^{-3}$ m, $\Delta t = 4 \times 10^{-4}$ s, $L = 0.5$ m, $T = 2$ s ($\Rightarrow M = 5001, K = 501$). The four

plots correspond to $D_r = 0.002$ (top left), $D_r = 0.02$ (top right), $D_r = 0.1$ (bottom left) and $D_r = 0.5$ (bottom right). In each plot, twenty equally spaced contours between $c = 0$ and $c = 1$ are shown, i.e., the contours correspond to curves of constant $c = m/19$, $m = 0, 1, \ldots, 19$.

## 5. Simulations of granular segregation: Variable kinetic stress profile

Fan and Hill [7] found, from particle dynamics simulations performed using the Discrete Element Method (DEM), that the kinetic stress profile for a dry bidisperse mixture in a vertical chute is well approximated by the expression $\sigma_{yy}^k(y) = A\exp(B|y|)$, where $A = 8.21 \times 10^{-4}$ N/m² and $B = 0.28$ mm$^{-1}$, while $\rho = 2500$ kg/m³, $q = 2.5 \times 10^{-3}$ s and $D = 0.05$ mm²/s.

A simulation of the segregation process under this model and parameters is shown in Fig. 4. Clearly, the initially well-mixed state (with $\phi_0 = 0.5$ here) quickly demixes near the boundaries, and these segregation fronts propagate into the middle of the domain. This is precisely what is observed in particle dynamics simulations using the Discrete Element Method [7]. The right plot in Fig. 4 below can be directly compared to Fig. 10 in [7]. Figure 10(b) in [7] gives a finite-difference approximation to same IBVP (1)–(2) as considered herein. Clearly, the scheme proposed in this work performs better (less numerical artefacts) on a coarser grid.

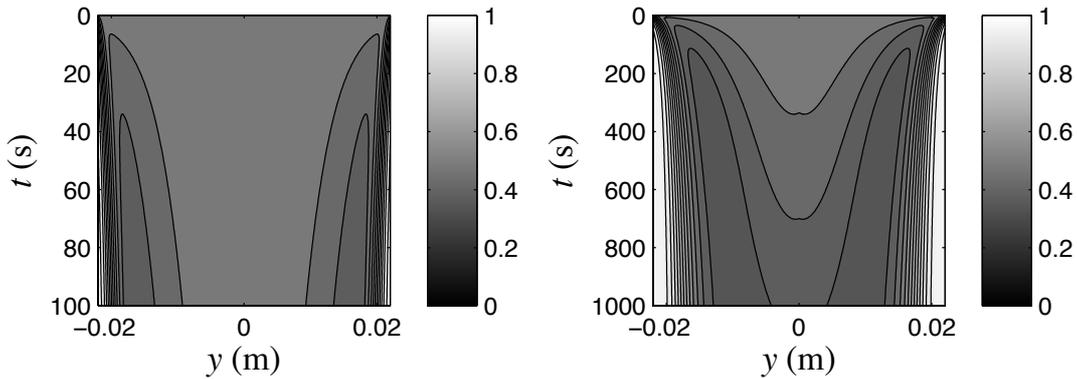

Figure 4. Contour plots of the spatiotemporal evolution of $\phi(y, t)$ (color bars denotes $\phi$ values for contours) as an initially uniform mixture segregates under a variable kinetic stress profile with $\Delta y = 8.8 \times 10^{-6}$ m, $\Delta t = 0.5$ s, $L = 0.022$ m, $T = 1000$ s ($\Rightarrow M = 2001, K = 501$). The left plot is over $t \in [0, 100\text{ s}]$, while the right plot is over $t \in [0, 1000\text{ s}]$.

## 6. Conclusion

In this short communication, we studied a variable-coefficient Burgers equation arising in the modelling of segregation of dry bidisperse granular mixtures. We constructed a strongly implicit, second-order accurate Crank–Nicolson-type scheme and benchmarked it against an exact kink-type solution. We applied this new scheme to two representative granular segregation problems, showing excellent agreement with results in the literature. Specifically, we match the exact results of Gray and Chugunov [9] and improve upon the explicit finite-difference scheme used by Fan and Hill [7,8]. Thus, we have demonstrated the utility and potential of the proposed numerical methodology. In fact, it is worth noting that we obtain excellent results on rather modest grids (a few thousand grid points), and the simulations take less than 10 seconds to run in MATLAB on a standard laptop computer.

Future work will include incorporating the possibility of an inhomogeneous diffusion coefficient $D = D(y, \dot{\gamma}, \ldots)$ [17]. The scheme presented above can easily be generalized to this case. Even a nonlinear diffusion coefficient $D = D(\phi)$ could be handled by this type of scheme with careful modifications, as shown in [18]. Additionally, it is of interest to compare these continuum solutions more carefully to those obtained by the DEM [3,5,7,8,17]. It may also be of interest to understand the Lie symmetries and group theoretic properties of the scheme that we proposed in Section 3 [19].

Finally, we note that continuum theory of granular mechanics, such as the model (1) studied herein, might be "heretic … [yet] also full of interesting possibilities" [20]. The present work focused on a purely kinematic model, however, *generalized* continuum mechanics (GCM) [21] may eventually shed further light on this interesting class of materials.

*Acknowledgements*

*The author would like to thank Prof. A.V. Porubov for his kind invitation to contribute to this special issue, and for the invitation to participate in the minisymposium "Nonlinear wave dynamics of generalized continua" (in memory of E. Aero and G. Maugin) at the XLV International Summer School – Conference "Advanced Problems in Mechanics — 2017". Thus, this manuscript is dedicated to the memories of Profs. E. Aero and G.A. Maugin. Last, but not least, the author would also like to express his gratitude to Dr. Yi Fan for many fruitful discussion (circa 2010-2011) on modelling granular segregation.*